\documentclass[fleqn,10pt]{wlscirep}
\usepackage[utf8]{inputenc}
\usepackage[T1]{fontenc}
\usepackage{lineno}
\usepackage{listings}
\usepackage{pifont}

\newcommand{\cmark}{\ding{51}}
\newcommand{\xmark}{\ding{55}}

\title{Rethinking the production and publication of machine-reusable expressions of research findings}

\author[1,2,*]{Markus Stocker}
\author[1]{Lauren Snyder}
\author[2]{Matthew Anfuso}
\author[2]{Oliver Ludwig}
\author[1]{Freya Thießen}
\author[1]{Kheir Eddine Farfar}
\author[1]{Muhammad Haris}
\author[1]{Allard Oelen}
\author[3]{Mohamad Yaser Jaradeh}
\affil[1]{TIB -- Leibniz Information Centre for Science and Technology, 30167 Hannover, Germany}
\affil[2]{Leibniz University Hannover, Institute of Data Science, 30167 Hannover, Germany}
\affil[3]{L3S Research Center, 30167 Hannover, Germany}

\affil[*]{corresponding author(s): Markus Stocker (markus.stocker@tib.eu)}

\begin{abstract}
Literature is the primary expression of scientific knowledge and an important source of research data. However, scientific knowledge expressed in narrative text documents is not inherently machine reusable. To facilitate knowledge reuse, e.g. for synthesis research, scientific knowledge must be extracted from articles and organized into databases post-publication. The high time costs and inaccuracies associated with completing these activities manually has driven the development of techniques that automate knowledge extraction. Tackling the problem with a different mindset, we propose a pre-publication approach, known as \emph{reborn}, that ensures scientific knowledge is born reusable, i.e. produced in a machine-reusable format during knowledge production. We implement the approach using the Open Research Knowledge Graph infrastructure for FAIR scientific knowledge organization. We test the approach with three use cases, and discuss the role of publishers and editors in scaling the approach. Our results suggest that the proposed approach is superior compared to classical manual and semi-automated post-publication extraction techniques in terms of knowledge richness and accuracy as well as technological simplicity. 
\end{abstract}

\begin{document}

\flushbottom
\maketitle

\thispagestyle{empty}

\section*{Introduction}

Articles have been the primary expression of research work and findings ever since Le Journal des sçavans \cite{desallo1665savans} and the Philosophical Transactions of the Royal Society \cite{oldenburg1665rstl} - two of the earliest published scientific journals - started operations in 1665. The fact that the article persists in the digital era of the 21st century arguably speaks for the usefulness of this expression in communicating research work from experts to experts.

However, when expressed solely in narrative text documents, scientific knowledge is not inherently machine reusable. In other words, it lacks formal (i.e. machine readable) syntax and semantics, preventing us from leveraging digital tools that could streamline research workflows \cite{shotton2009wiley}. Indeed, published scientific knowledge is routinely ``buried'' in documents \cite{mons2005bmc} and currently relies (largely) on manual knowledge extraction activities \cite{higgins2023cochrane} to make it ready for machine-supported processing. For instance, conducting synthesis research in the form of a meta-analysis or systematic review requires manually extracting data from tens or sometimes hundreds of text-based articles and organizing this data into a new database, which is a time-consuming and error-prone process. One study estimated that completing a systematic review takes between six to 12 months of full-time work, with roughly half of this time devoted to pre-analysis literature search, data retrieval, and database development \cite{allen1999jama}.

To facilitate the efficient transfer, reuse, and synthesis of scientific knowledge, there is a growing movement to ensure knowledge expressed in scientific articles is produced in a manner that is findable, accessible, interoperable, and reusable (FAIR) for humans, and also for machines \cite{wilkinson2016nsd}. Creating machine-reusable expressions of scientific knowledge plays a key role in FAIR data practices as it improves the capacity of machines to autonomously discover and utilize scientific knowledge, thereby facilitating its reuse by human researchers \cite{attwood2009bio,auer2020biblio}. Yet, while organizing scientific knowledge (only) as a collection of articles has been challenged for some time and the development of approaches for more advanced scientific knowledge organization has received considerable attention (\cite{hars2001inf,waard2009swasd,groth2010inf,shotton2009plos,iorio2015semantic}), the systematic production of machine-reusable scientific knowledge remains elusive.

An obvious class of approaches centers on extracting knowledge from published articles, which can occur manually (by human experts) or semi-automatically (with computer assistance). Manual approaches to producing machine-reusable scientific knowledge can be supported by specialized user interfaces \cite{kuhn2021peerj,jaradeh2019kcap}. While high-quality data can be achieved, manual production hardly scales. Given the appeal of automation, the application of Natural Language Processing (NLP), Machine Learning (ML), Text Mining (TM), or similar techniques that allow machines to process human language in narrative text articles for information extraction continues to receive substantial attention in the literature \cite{agrawal2019workshop,hong2021jom,groth2018comp,nasar2018scie,salloum2017intel}, fueled also by the recent developments in generative AI and Large Language Models (LMM) \cite{dagdelen2024nature}. True automation has, however, not been achieved. Indeed, as the following task suggests, with a performance of 10\% \cite{hou2019comp} the extraction of simple TDMS (Task, Dataset, Metric, Score) tuples from articles in ML cannot be automated. More generally, scientific knowledge extraction tasks are considerably more challenging than the TDMS task because scientific knowledge typically has a more complex structure and the presentation of scalar values in tables typical for TDMS reporting in articles is a special case; more generally, data presented in articles are matrices and are often displayed in figures. 

In contrast to post-publication methods for knowledge extraction, approaches that center on producing machine-reusable scientific knowledge pre-publication, i.e. before the related manuscript(s) is/are published, have been less pursued, but could offer a number of advantages to researchers, editors, and publishers. For instance, SciKGTeX \cite{bless2023jcdl} is an example of a system that integrates with \LaTeX~manuscript editing environments to support the production of machine-reusable scientific knowledge through annotations of narrative text documents. As an additional example, dokieli \cite{capadisli2017lncs} is a document authoring environment with built-in support for annotations encoded as structured data.

Here, we use a series of use cases to present and evaluate a novel pre-publication approach, known as \emph{reborn}, that integrates with computing environments for (statistical) data analysis, ensuring scientific knowledge is born-reusable (i.e. expressed in a machine-reusable format upon production). Grounding in earlier phases of the research lifecycle, we propose an end-to-end distributed system that spans machine-reusable scientific knowledge production, deposition/publishing, and collection in aggregation systems that support efficient knowledge reuse. Thus, the present work addresses the following research question: How can we support the production of machine-reusable expressions of research findings in the data analysis phase of the research lifecycle?

Specifically, we propose an interoperable distributed system that enables the pre-publication production of machine-reusable scientific knowledge in (statistical) computing environments (e.g. RStudio or Jupyter and the languages R and Python); the publishing of such expressions as supplementary data interlinked with articles in publisher digital libraries or data repositories (e.g. Zenodo, Dryad); and the automated discovery and collection of such content, and content reuse (e.g. for synthesis research), for which we leverage the Open Research Knowledge Graph \cite{stocker2023fc} (ORKG), a digital scholarship infrastructure that supports the production, curation, and reuse of machine-reusable scientific knowledge as FAIR research data. We present \emph{reborn} in three use cases involving three articles published in journals or conference proceedings by different publishers and demonstrate that this approach is ready for immediate implementation by authors. 

We discuss the role of publishers and journal editors in scaling the approach, contrast the approach with classical knowledge extraction, and discuss possible implications for the review process, science reproducibility and transparency, and synthesis research. While we evaluate \emph{reborn} primarily on quantitative (statistical) scientific knowledge, we suggest that it is broadly applicable to arbitrary (scientific) knowledge types for which it is possible to develop a schema. Indeed, by applying \emph{reborn} to the present article (see data and code availability statement), we provide a use case demonstrating its application to qualitative knowledge. Finally, we also remark that while the approach is conceived for future research, the use cases show that it can also be applied retroactively to already published articles.

\section*{Results}

\begin{figure}[!ht]
\centering
\includegraphics[width=\linewidth]{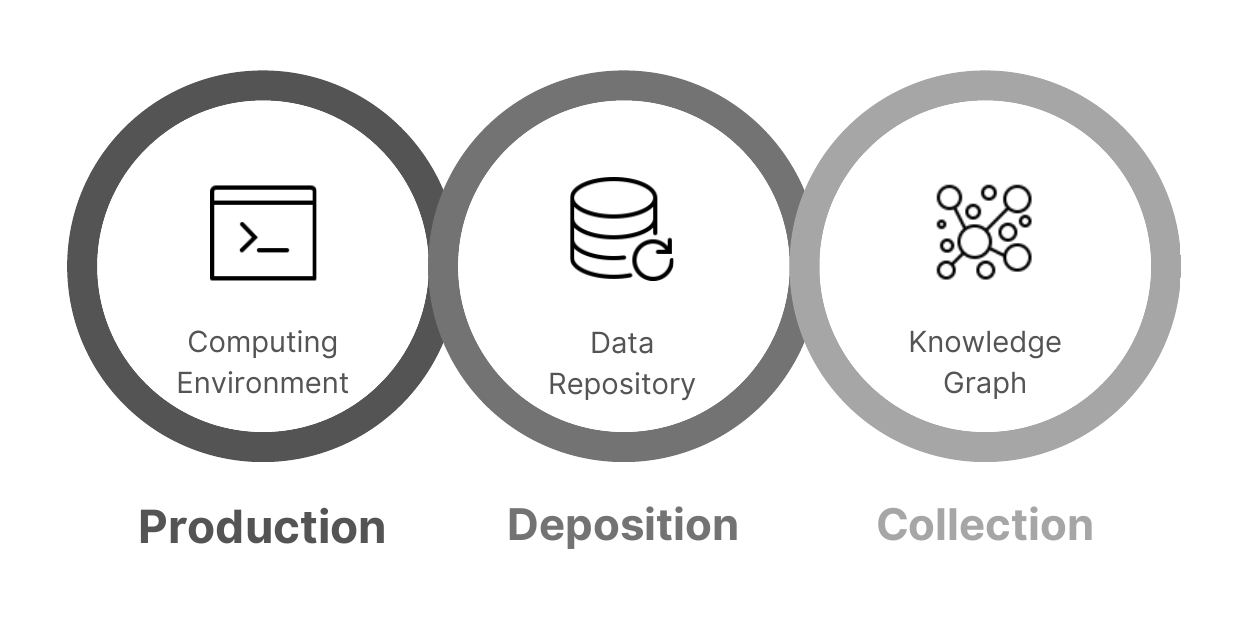}
\caption{Scientific knowledge expressed in articles is produced as machine-reusable data  in computing environments during the data analysis phase of the research lifecycle. Machine-reusable scientific knowledge is deposited in a data repository as supplementary data of the article and interlinked with the article in DOI metadata. Finally, to support reuse, e.g. for synthesis research, machine-reusable scientific knowledge is collected and organized in aggregation systems, such as knowledge graphs.}
\label{fig:figure1}
\end{figure}

Figure \ref{fig:figure1} provides a high-level overview of the proposed approach. It consists of three stages: (1) The production of machine-reusable scientific knowledge in computing environments during the data analysis phase of the research lifecycle; (2) the deposition of machine-reusable scientific knowledge as supplementary data interlinked with the published article; (3) the collection of deposited machine-reusable scientific knowledge in aggregation systems such as knowledge graphs \cite{hogan2021acm}. In this section, we describe \emph{reborn} along these stages and how we applied the approach in three use cases. A more detailed technical account of the implementation is provided below in the Methods section. 

\paragraph{Production.} Scientific knowledge expressed in articles, also known as final data, is often produced in the (statistical) data analysis phase of the research lifecycle. With \emph{reborn}, we ensure that scientific knowledge is also produced machine reusable, i.e. scientific knowledge is born reusable. This is achieved by extending the implementation of data analysis with additional instructions that implement the production of machine-reusable final data, i.e. machine-reusable expressions of knowledge production processes, including their input and output data items. At the core of these additional instructions is the integration of data type schemata that guide researchers in describing their findings in a structured manner (further details provided in the Methods section). This promotes a standard representation of research findings and enhances the comparability of findings across articles. The resulting machine-reusable expressions are then conceptualized as supplementary data of the manuscript prepared later in the research lifecycle, where data items are presented as results as well as materials and methods.

For instance, as a doctoral student, Darya uses RStudio to perform a Student’s t-test to compare the means of control and treatment groups for a dependent variable of interest. She reports her results as plotted observations and a p-value in her manuscript. In Darya’s work, the plotted observations and the p-value (i.e. the input and output data, respectively, in the Student’s t-test) are final data. Being familiar with \emph{reborn}, Darya extends her R script with the required additional instructions to describe the conducted Student’s t-test as a machine-reusable expression of the research finding (see Listing \ref{lst:listing1} in the Methods section). These additional instructions in her R script produce supplementary data.

\paragraph{Deposition.} Upon manuscript finalization, Darya submits the supplementary data expressing machine-reusable scientific knowledge together with her manuscript to the journal or conference of choice. Alternatively, Darya may also deposit the supplementary data in the preferred data repository (e.g. Zenodo or comparable). Either approach ensures the supplementary data enters the review and publication workflow. Journal editors and conference chairs may make the supplementary data available to reviewers to support the review process, including verifying the correctness of final data reported in the manuscript.

Upon article publication, the publisher ensures permanent (open) access to the supplementary data as well as data discoverability. Discoverability (for machines) can be supported by interlinking supplementary data and the article in DOI metadata. With the so-called ``related identifiers'' metadata attribute, metadata schemata by Crossref and DataCite already implement an appropriate mechanism for such interlinking. Given the widespread implementation of automated link information exchange among scholarly infrastructures, such links are also easily discoverable in systems such as DataCite Commons (\href{https://commons.datacite.org}{commons.datacite.org}), OpenAIRE Research Graph (\href{https://graph.openaire.eu/}{graph.openaire.eu}), or similar.

\paragraph{Collection.} Given the article's DOI, aggregation systems can discover and collect machine-reusable scientific knowledge. In our work, we utilize the ORKG as an aggregation system. A key role of such systems is the provision of value-added services, in particular for efficient access, processing, and visualization of scientific knowledge required, e.g. for synthesis research. The interlinking of the original article with the corresponding machine-reusable expression of the research findings published in the article is a simple example for such a service. As the present work focuses on the production and publication of machine-reusable scientific knowledge, a more detailed exploration of value-added services for knowledge reuse is not the focus of this article.

\subsection*{Use cases}

We now present how we tested \emph{reborn} in three use cases involving articles published in journals or proceedings by different publishers. For each use case, we describe the results along the three stages. Further technical details are provided in the Methods section using the first use case in soil science as a running example.

\subsubsection*{Use case in soil science}

The use case in soil science centers around the paper by Gentsch et al. (2023) \cite{gentsch2024soil} titled ``Cover crops improve soil structure and change organic carbon distribution in macroaggregate fractions'' published in the journal SOIL by Copernicus Publications (\href{https://doi.org/10.5194/soil-10-139-2024}{doi:10.5194/soil-10-139-2024}). Figure 2 in this article illustrates the result of executing the proposed approach for this use case. For an interactive experience, we refer readers to the version published at \href{https://doi.org/10.48366/R664252}{doi:10.48366/R664252}.

\paragraph{Production.} Gentsch et al. implemented their data analysis in R. We focus the use case on the main research findings, presented in Figure 1, Table 1, Figure 2 a) and b), and Figure 3 of the original article by Gentsch et al. These elements each result in some form of statistical computation. Figure 1 presents descriptive statistics using radar charts. The data presented in Table 1 and Figure 2 b) are a result of linear mixed effects model (LMM) fittings. Figure 2 a) is a result of pairwise t-tests. Finally, Figure 3 is a result of structural equation modeling. For each of these statistical methods, we created ORKG Templates (i.e. schemata that specify data structures; for details, we refer readers to the Methods section) to support the description of data analysis activities, in particular input data and implementation as R script snippet, as well as output data, visualization, a human-readable statement, and a (statistical) model description (if applicable). We then integrated these templates with the original R scripts to support the programmatic production of machine-reusable expressions of the research findings as supplementary data. The supplementary data were produced in close collaboration with Norman Gentsch, the first author of the work, before manuscript submission to the journal. 

\paragraph{Deposition and Collection.} To streamline the process and ensure a quality publication, including the reproducibility of research findings, we implemented data deposition ourselves (\href{https://doi.org/10.57702/yztrbsd4}{doi:10.57702/yztrbsd4}). In addition to depositing the supplementary material, we interlinked the data deposition with the article in data DOI metadata. Given the article DOI, aggregating systems such as the ORKG can discover the interlinked supplementary material and collect the published machine-reusable scientific knowledge, organize it in databases, and make knowledge available for efficient reuse.

\begin{figure}[!ht]
\centering
\includegraphics[width=\linewidth]{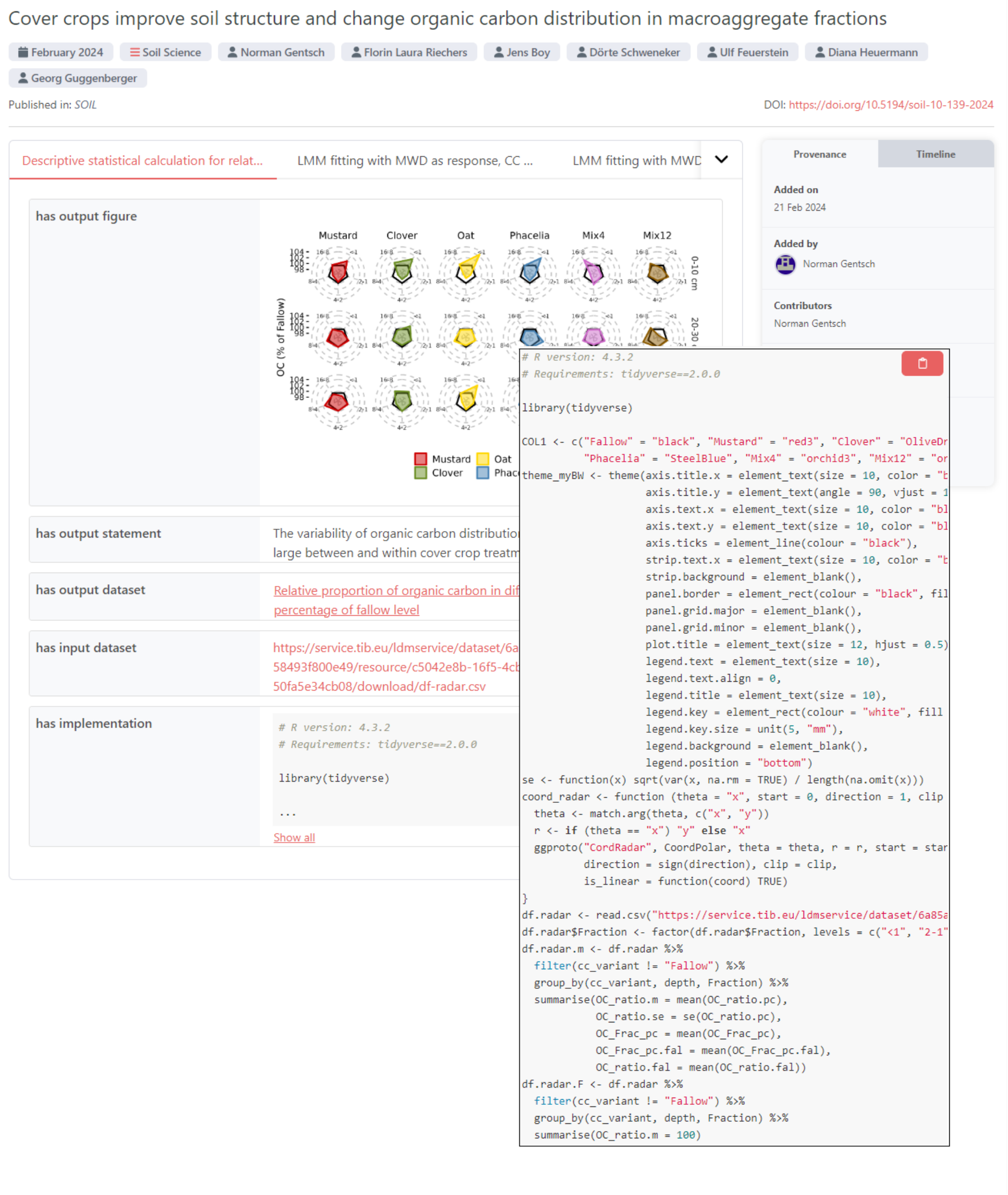}
\caption{Display of the research finding published by Gentsch et al. in their Figure 1 as a research contribution in ORKG. The overlay expands on the interlinked R script snippet used to implement the respective data analysis. For an interactive experience, we refer readers to the version published online at \protect\href{https://doi.org/10.48366/R664252}{doi:10.48366/R664252}.}
\label{fig:figure2}
\end{figure}

\paragraph{Highlight.} In this first use case, we highlight the potential of interlinking script snippets used to produce research findings, described in ORKG as research contributions. Figure \ref{fig:figure2} expands on such a script snippet for the research finding presented in Figure 1 of the original article by Gentsch et al. Assuming the input data required for the analysis are published and accessible online, we can interlink the input data in the script to support the frictionless reuse of data and code, for instance to easily reproduce and verify the correctness of the published results. Such possibilities could be enabled specifically for reviewers during the manuscript review phase, but are useful to research more broadly, in particular for synthesis research. Furthermore, in collaboration with Copernicus Publications, we achieved the interlinking of the machine-reusable expressions of the research findings by Gentsch et al. in the ORKG as a data set asset of the original article, a relation that human experts can discover on the article landing page (\href{https://soil.copernicus.org/articles/10/139/2024/soil-10-139-2024-assets.html}{soil.copernicus.org/articles/10/139/2024/soil-10-139-2024-assets.html}) and machines in Crossref DOI metadata (\href{https://api.crossref.org/works/10.5194/soil-10-139-2024}{api.crossref.org/works/10.5194/soil-10-139-2024}).

\subsubsection*{Use case in computer science}

The use case in computer science centers around the paper by Thießen et al. (2023) \cite{thiessen23ceur} titled ``Probing Large Language Models for Scientific Synonyms'' published in the proceedings of the 2nd NLP4KGC workshop by CEUR Workshop Proceedings. This work falls into the broad category of research in Machine Learning (ML) whereby the goal is to develop and evaluate the performance of models for some ML task using suitable benchmark datasets and metrics. The main research findings are thus the performance scores and the most salient data are for so-called TDMS-tuples (Task, Dataset, Metric, Score). In the work by Thießen et al., the Task is ``Synonym Discovery'' and the developed models are evaluated using a number of benchmark Datasets and Metrics. The machine-reusable expression of the published research findings is available at \href{https://doi.org/10.48366/R661500}{doi:10.48366/R661500}.

\paragraph{Production.} Based on several evaluation datasets, the research findings by Thießen et al. consist of F1 scores for the performance of different Large Language Models in identifying synonyms. The production of machine-reusable expressions of TDMS-tuples was supported by integrating the ORKG Leaderboard template in the Python script that implemented data analysis. The Leaderbord template captures TDMS-tuples in a schema that, in the ORKG, is shared across articles that also evaluate the performance of models for some ML task. In this use case, the article was submitted and published before we created the supplementary data as presented here. 

\paragraph{Deposition and Collection.} We use the same deposition and collection approach already described in the first use case. The supplementary data deposition is published at \href{https://doi.org/10.57702/9zhuubz9}{doi:10.57702/9zhuubz9}. Since the article was published by CEUR-WS, we were unable to interlink supplementary data with the article in DOI metadata. This is because CEUR-WS does not identify published articles with DOI; the articles are merely linked by their URL on a landing page for the volume representing the workshop proceedings. Given these constraints, we used a file-based, rather than an article DOI-based, collection in ORKG (see the Methods section for more details).

\begin{figure}[!ht]
\centering
\includegraphics[width=\linewidth]{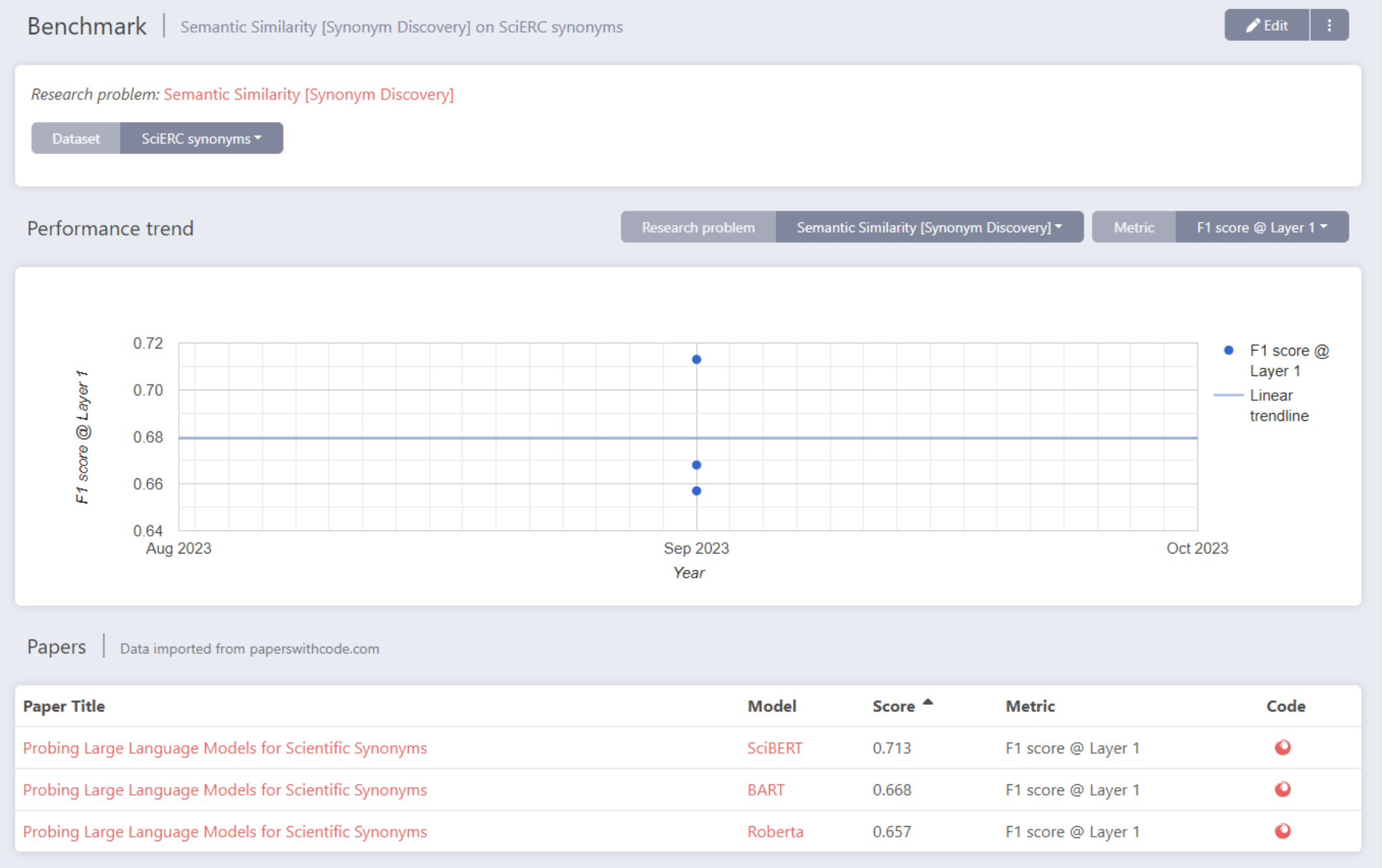}
\caption{Display of a Leaderboard showing the performance Scores (Metric F1 score @ Layer 1) of the three models evaluated using the SciERC Dataset for the machine learning Task of ``Synonym Discovery'' as published by Thießen et al.}
\label{fig:figure3}
\end{figure}

\paragraph{Highlight.} An interesting aspect of this type of research is that TDMS-tuples aggregate to so-called Leaderboards frequently found in ML research communities to visualize the state of the art in model development for a given ML task as well as model performance trends over time for the task. Papers with Code (\href{https://paperswithcode.com}{paperswithcode.com}) is arguably the most well-known aggregator system for such scientific knowledge and a pioneer in developing services that enable access to the state of the art in a large research community. In ORKG, data structured according to the ORKG Leaderboard template are automatically used in the construction of such Leaderboards. For the ML task ``Synonym Discovery'' and the SciERC dataset, Figure \ref{fig:figure3} plots the performance in terms of applicable metric and score of models as they were published by Thießen et al. in September 2023. This is an example of a value-added service by an aggregation system, here the ORKG automatically producing a derivative data product for TDMS data with common task.

\subsubsection*{Use case in agroecology}

\begin{figure}[!ht]
\centering
\includegraphics[height=\linewidth]{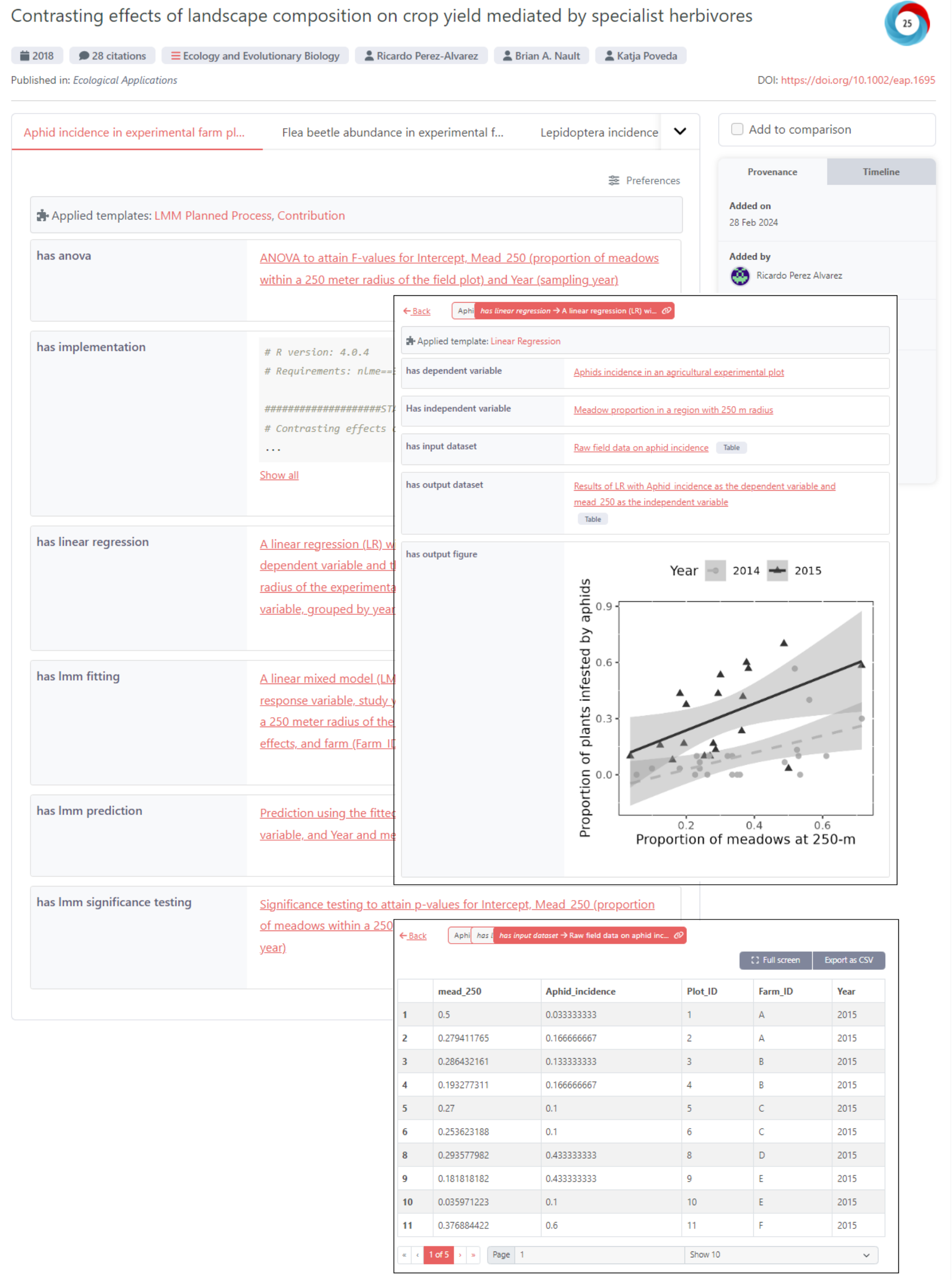}
\caption{Display of the research finding published by Perez-Alvarez et al. in their Figure 4 (a) as a research contribution in ORKG. The two overlays illustrate detailed information in the form of visualizations and tabular data. For an interactive experience, we refer readers to the version published online at \protect\href{https://doi.org/10.48366/R689181}{doi:10.48366/R689181}.}
\label{fig:figure4}
\end{figure}

The use case in agroecology centers around the paper by Perez-Alvarez et al. (2018) \cite{perez2018wiley} titled ``Contrasting effects of landscape composition on crop yield mediated by specialist herbivores'' published in the journal Ecological Applications by Wiley (\href{https://doi.org/10.1002/eap.1695}{doi:10.1002/eap.1695}). Figure \ref{fig:figure4} in this article illustrates the result of executing the proposed approach for this use case. For an interactive experience, we refer readers to the version published online at \href{https://doi.org/10.48366/R689181}{doi:10.48366/R689181}. We present this third use case in a succinct manner in this paper as it will be described in further detail in a forthcoming manuscript by Snyder et al.

Perez-Alvarez et al. implemented their data analysis in R and used linear mixed effects model (LMM) fittings, among other statistical methods. As the article by Perez-Alvarez et al. was published in 2018, we achieved the retroactive production and deposition of the supplementary data in close collaboration with Ricardo Perez-Alvarez, the first author of the work. As with the first use case, metadata interlinking enabled the ORKG to automatically discover and collect supplementary data (\href{https://doi.org/10.57702/eyff90ic}{doi:10.57702/eyff90ic}) given the DOI of the article by Perez-Alvarez et al. The result of this stage is illustrated in Figure \ref{fig:figure4}.

\paragraph{Highlight.} In this third use case, we highlight how \emph{reborn} supports expressing rich scientific knowledge, including complex tabular data, in machine-reusable form.

\section*{Discussion}

As its primary goal, the present work proposes an approach that supports the production of machine-reusable expressions of research findings in the data analysis phase of the research lifecycle. Additionally, the work presents how these expressions can enter the manuscript submission, review, and article publication phases as interlinked supplementary data, collectable by aggregation systems that support their reuse in research. Having presented a high-level overview of the \emph{reborn} approach, we evaluated its practical viability in three use cases. In the Methods section, we describe our reference implementation for the proposed approach and the distributed system along its main supported activities.

Our work is motivated by the fact that the traditional expression of research work as narrative text documents is hindering the efficient use of the scientific knowledge expressed in the scholarly record. This is particularly evident in synthesis research, where researchers routinely manually extract relevant data from articles, an activity that is not only time consuming, but often also inaccurate. More generally speaking, the scholarly infrastructure supports finding and accessing documents, but is unable to further assist researcher information needs and information processing. As a result, a researcher in machine learning can find articles possibly relevant to their research in image classification, but the infrastructure is unable to inform which model currently represents the state of the art.

\subsection*{Comparing approaches}

The systematic production of rich and accurate machine-reusable expressions of scientific knowledge at scale presents a formidable obstacle towards FAIR scientific knowledge. In Table \ref{tbl:table1}, we compare manual and automated post-publication extraction of scientific knowledge with the proposed pre-publication production approach along several dimensions. By excelling in accuracy and richness, the overview makes evident that \emph{reborn} addresses important dimensions that remain a challenge for post-publication extraction. Moreover, the overview also suggests that the proposed approach is competitive in several other dimensions.

\paragraph{Accuracy.} In terms of accuracy, the extraction of scientific knowledge from published articles continues to challenge automation. Automated approaches can achieve above 90-95\% accuracy, but only for the simplest of tasks \cite{jonnalagadda2015sys}. For example, authors report an accuracy of $\sim$90\% for the extraction of TDM (Task, Dataset, Metric, \emph{without} Score) tuples from papers in machine learning \cite{kabongo2021lncs}. As another example, using a well-engineered set of prompts, a new conversational LLM approach to post-publication data extraction achieved 91.6\% accuracy extracting materials data as (Material, Unit, Value) tuples from articles \cite{polak2024nature}. In a further experiment in materials science, the authors concluded that even advanced LLMs struggle to extract all required information from articles \cite{khalighinejad2024arxiv}. With performances of approximately 60\%, 40\%, and 30\% accuracy, respectively, the more general tasks of keyphrase identification, keyphrase classification, and relation extraction between keyphrases further underscore the difficulty of automating increasingly difficult tasks \cite{augenstein2017comp}. In contrast to automated post-publication extraction, with manual post-publication extraction it is possible to capture more detailed information with higher accuracy, as human experts can perform complicated data extraction tasks with a higher accuracy than machines. The chance of human error in manually copying or transforming data is, however, not negligible \cite{mathes2017bmc,gotzsche2007jama}. Post-publication extraction also suffers from the problem that data in articles may be approximate. Examples include an approximate p-value or plotted data where data extraction tools are used to estimate the underlying quantitative values. In contrast, pre-publication production as proposed in the present work also benefits from direct access to input and output data, say the CSV data and p-value that are input and output to a statistical hypothesis test. By embedding with data analysis, \emph{reborn} also avoids copying errors as it can leverage the passing of variables in scripts.

\paragraph{Richness or granularly.} This dimension is particularly interesting to contrast between post-publication extraction and pre-publication production. Both manual and automated approaches struggle to extract rich or highly granular, machine-reusable expressions of scientific knowledge from articles. Manually, the task simply requires too much effort from individual researchers. Moreover, we argue that there is currently little or no evidence that collaboration may alleviate the problem. While crowd-based collaboration has succeeded in some projects, for instance for encyclopedic knowledge (e.g. Wikipedia) or spatial information (e.g. OpenStreetMaps), to the best of our knowledge, there is currently no evidence that crowd-based collaboration has succeeded or will succeed for scientific knowledge. As the works cited earlier suggest, automated approaches currently similarly fail to extract rich, highly granular, knowledge from articles. The main hindrance is the overall complexity of the task as scientific knowledge is, in general, a complex data type involving numerous, highly-related entities. To make matters worse, related data are often presented in text as well as figures, tables, listings, or similar. In contrast, by embedding expressive data type schemata into data analysis, \emph{reborn} enables the relatively effortless production of rich machine-reusable expressions of scientific knowledge.

\paragraph{Technological simplicity.} In comparison, post-publication automated extraction is by far the most technologically complex class of approaches. This is because algorithmic information extraction from articles relies on quality training data, the construction and evaluation of models, as well as model deployment into production infrastructure. Moreover, these activities are often or even mostly task specific: The extraction of a quantity value with confidence interval may be best implemented using rules (regular expressions) while the extraction of TDMS-tuples (Task, Dataset, Metric, Score) from articles in machine learning may be best implemented using NER (Named-Entity Recognition) or a language model. In contrast, both post-publication manual extraction and pre-publication production are technologically by far more feasible, as they rely primarily on software development (e.g. user-friendly frontend or software libraries). This is a clear, but sometimes neglected, advantage for the practical viability of a system.

\paragraph{Scalability.} Assuming a very good algorithmic performance, the ability to efficiently scale to millions of articles is, unsurprisingly, a characteristic of automated approaches. Relying on some level of manual activity, both post-publication manual extraction and pre-publication production inherently underperform on this aspect. The fact that these classes of approaches also rely on awareness and training in research communities further complicates their scalability.

\paragraph{Coverage.} Given automation, the potential to cover the literature of individual research communities, in disciplines or in research more generally, is highest with post-publication automated extraction. Again, very good algorithmic performance is a precondition. However, assuming broad adoption and participation, it is arguably possible to achieve good coverage also with manual post-publication extraction and pre-publication production. This is particularly the case for specific research communities or research fields, which is made evident by numerous projects including Papers with Code in Machine Learning (\href{https://paperswithcode.com}{paperswithcode.com}), Hi Knowledge in Invasion Biology (\href{https://hi-knowledge.org/}{hi-knowledge.org}), and Plazi in Biodiversity (\href{https://plazi.org/}{plazi.org}). 

\paragraph{Legacy.} The ability to consider already published articles, decades or even centuries ago, is a characteristic of post-publication extraction. Here, too, automation surely helps scaling to the millions of articles that have been published over the past centuries. However, on born-analogue, scanned articles, an OCR (Optical Character Recognition) step further complicates post-publication extraction. Manual extraction can be easily performed on long published articles, too. In contrast, pre-publication production is a class of approaches that relies on integration in the research lifecycle and is thus designed for future research. However, assuming data and scripts are interlinked with already published articles, \emph{reborn} can in principle be performed also post-publication. This was indeed the case for our second and third use cases presented earlier. Unfortunately, the systematic publishing of data and scripts as supplementary data to articles is not yet common practice, which severely limits the retroactive application of our approach.

\subsection*{Role of publishers and journals}

As we suggested above, \emph{reborn} is technologically simple, especially if compared to automated post-publication extraction (Table \ref{tbl:table1}). However, scalability is a challenge. Publishers and journals can play a major role in addressing scalability, and can do so in numerous ways. First, journals can guide prospective authors to implement the approach via their Author Guidelines. The approach's essence is straightforward, as it amounts to incorporating an additional library into a script, writing a few additional lines of code that describe the (key) results, capturing the additional data produced in the process, and submitting this data together with the manuscript to the journal of choice. Hence, the guidance to authors by journals can be limited to a short description. With growing adoption, it will also be possible for journals to point to examples of published papers with related machine-reusable expressions of research findings.

To ensure researchers can easily apply \emph{reborn} to create machine-reusable expressions of findings for a range of research methodologies and statistical analyses, it will be important to have a representative registry of ORKG Templates (more generally, data type schemata that specify research findings). Ideally, the creation of templates is a process driven by community consensus. Therefore, a key step will be to establish an efficient process that ensures the creation of templates in response to a research community’s need. One approach is to develop so-called ORKG Observatories, where a group of experts in some research domain join forces to identify knowledge types relevant to their community and specify such types as templates. While this remains untested, Observatories could become a viable approach to efficiently serve communities with the required templates. 

One advantage of templates is that they are often highly reusable across disciplines; for instance, the Student's t-test is a popular statistical method used in natural sciences, life sciences, as well as in social sciences and the humanities. Once developed by one research community, such a template could be applied across multiple disciplines. The Statistics Ontology (STATO, \href{https://stato-ontology.org}{stato-ontology.org}) classifies statistical methods such as types of statistical hypothesis tests or regression analyses as data transformations. In total, STATO defines about 180 data transformation types. It is plausible that their usage in research follows a power law. Therefore, in addition to being reusable, we argue that a relatively small set of templates could achieve a high coverage of published research findings. As a result, the impact of such a registry would easily justify the investment required to specify the widely needed templates.

Publishers also have a more technical role in the process. The supplementary data for the machine-reusable expressions of research findings as well as related materials (e.g. scripts, images, or associated datasets) submitted together with the manuscript need to be individually published and persistently accessible at some URL. To enable discovery of such supplementary data, it is useful to interlink these assets as related identifiers in DOI metadata. While authors may publish supplementary data in a data repository of choice (e.g. Zenodo, Dryad, etc.), interlinking supplementary data in article DOI metadata is a task that relies on publishers.

\subsection*{Implications for the review process}

The proposed approach aims to support the more efficient reuse of scientific knowledge in synthesis research. However, the supplementary data submitted together with manuscripts to journals can also strengthen the peer-review process. Indeed, reviewers of manuscripts may derive several benefits from structured expressions of research findings. 

First, the more granular structuring may increase clarity compared to the more compact presentation of the same information in manuscripts. Second, the interlinking of terms with terminologies can also increase clarity, particularly in fields like ecology where terminology is frequently redefined across disciplines, foundational terms are commonly ignored, and synonymous terms abound \cite{herrando2014bio}. Third, by reusing structures specified by templates, the machine-reusable expressions of scientific knowledge may be more standardized across the literature than the corresponding expressions in articles, where authors do not necessarily agree on using some standard vocabulary. Greater standardization could also support the review process. Fourth, the granular interlinking of data can support verifying the correctness of data visualized as figures in manuscripts. Fifth, the interlinking of scripts can support validating the published results as well as their reproducibility. Though AI can generate data and code in addition to narrative text \cite{naddaf2023nature}, structured and reproducible expressions of research findings as proposed here may, as a sixth benefit, support detecting fabricated research, especially if final data are also related to primary data and contextual information about the research lifecycle (e.g. instruments used to collect primary data) in provenance metadata.

For these reasons, we suggest that it may be useful to make the submitted supplementary data available to reviewers during the review phase. Since at this point the results are not published, we suggest this should occur under embargo, accessible for reviewers only. By applying \emph{reborn} to the present article (see data and code availability statement), we enable reviewers (and readers) of this work to test some of our claims on the implications for the review process.

\subsection*{Impact}

Before we briefly reflect on the potential impact of \emph{reborn}, we underscore that its greatest potential impact applies primarily to research to be conducted in the future, and is limited for research conducted in the past. This focus on future research is, however, not a strong limitation. The production of scientific knowledge and its publication in scholarly literature has increased ever since the first journals began publishing scientific findings in the 17th century. Since 1950, science has grown exponentially, with a doubling time of 14 years \cite{bornmann2021hum}. This shows that while millions of articles have been published, only a few decades are (likely) required for future research to be more voluminous than past research. Furthermore, as the state of the art is of primary relevance to future research, past research arguably has a tendency to depreciate. Hence, any time is a good time to begin with disruptive approaches; if we implement more efficient pre-publication approaches now, we predict reliance on post-publication extraction will likely decrease in the coming decades, and possibly become largely obsolete in the second half of the century.

Given its technical simplicity and wide applicability to (at least) statistical scientific knowledge, we argue that the most important factor determining future impact of the approach is the readiness of researchers to adopt it in their research. The readiness of researchers to publish data and scripts related to articles is a related practice with lower requirements that would enable the proposed approach and is nowadays widely considered to be best practice for open and reproducible science and state-of-the-art research data management. As awareness and implementation of such practices percolates through disciplines and research communities, we expect an increase in articles that are interlinked with data and scripts as supplementary data. This is encouraging because such practices lay the foundations for implementing \emph{reborn}, as anyone with access to the article and the supporting data could in principle apply the proposed approach to create machine-reusable expressions of the published scientific knowledge, independently from the original authors.

Adoption of the presented approach will arguably be slower than adopting the practice of publishing supporting data and scripts related to articles. This is because, compared to simply collecting related files and submitting them as a package, the proposed approach comes with an overhead cost for researchers, especially in the form of time required to master and implement the technique. A key next step is to explicitly measure the overhead costs associated with \emph{reborn} and support its continued development to streamline its implementation across scientific disciplines.

However, embedding the production of machine-reusable expressions of research findings in data analysis may have important advantages, also in terms of efficiency. Indeed, \emph{reborn} is, broadly speaking, an important contribution to ensuring FAIR research data. Computing environments and languages such as Python and R harbor considerable potential for ensuring research data are FAIR, and this potential surely remains underexploited. The proposed approach is an excellent example demonstrating the potential of ensuring research data are created FAIR at production, by design of infrastructure, leveraging the power of computing environments and the established publication procedures in scholarly communication.

Hence, as research communities continue to strive towards best practices in open and reproducible science, we expect that \emph{reborn} or comparable approaches will gain traction and be important drivers in the production of machine-reusable expressions of research findings. As a result, the share of data produced pre-publication will increase and may ultimately overtake the share of data produced via post-publication manual or automated extraction, especially high-quality data reflecting rich machine-reusable expressions of research findings.

A quick analysis of approximately 90K Zenodo records showed that about 5K ($\sim$5\%) contain either Python or R files and are interlinked (in metadata) to about 8K published articles. While 8K articles are a very small fraction of the millions of articles published each year \cite{johnson2018stm}, the Zenodo data does suggest that in the course of a few years, hundreds, and in the longer run thousands, of articles could be interlinked with machine-reusable expressions of the research findings published therein, an important impact and impulse for the future of digital scholarship and the efficient reuse of scientific knowledge as FAIR research data.

\subsection*{Limitations}

The presented work and the proposed approach have a few limitations, which we discuss in more detail. First, the proposed approach integrates with data analysis conducted in statistical computing environments. While much of scientific knowledge is quantitative and statistical, not all scientific knowledge is. The approach has so far not been tested on qualitative or other kinds of scientific knowledge, e.g. formulae, proofs, arguments, etc. While untested, we suggest that assuming a knowledge type can be structured and specified with a template, then \emph{reborn} should be transferable, possibly with adaptations (e.g. if a certain kind of knowledge is not produced in a computing environment).

Second, we developed and evaluated the approach for R as a statistical computing language, and support the same also for Python. It is arguably possible to extend to other open or proprietary scripting languages, such as Julia or Matlab. More challenging is integrating the approach with statistical computing environments that lack scripting support or that are primarily used via a graphical user interface. Even more challenging are commercial solutions, such as SPSS, GraphPad Prism, Stata, which, according to some studies \cite{masuadi2021cureus}, continue to be not only top ranked but also dominate the market. Given reliance on software companies, integration in commercial applications seems a remote possibility from today’s vantage point. However, there is little reason to believe that such an integration is technologically considerably harder compared to integration in open statistical computing environments with support for scripting.

Third, our evaluation is limited, both in number of cases as well as in diversity of research. While we aim to address this limitation in future work, the main objective of the present work is to introduce the approach and provide evidence for its viability. However, based on the provided evidence, it seems promising that the proposed approach can be applied to scientific knowledge produced in statistical data analysis in research more generally.

A further limitation is the ORKG-specific syntax of the data produced using ORKG libraries and templates in Python- or R-based statistical computing environments. This is due to the fact that the templates integrated into scripts are created in ORKG and, naturally, use ORKG terminology. Furthermore, ORKG libraries produce JSON-LD formatted data that can directly be consumed by ORKG. This tight integration with ORKG has simplified the development of the \emph{reborn} approach at this stage. However, since the data are produced and published in a distributed manner and can in principle be consumed by systems other than the ORKG, we suggest that the published machine-reusable expressions of research findings should conform to a more widely accepted syntax and terminology. We plan to address this limitation in future work and provide some initial thoughts on a possible implementation in the corresponding section below.

\subsection*{Future work}

We applied the approach to three articles published by three publishers. In future work, we aim to scale the number of articles and journals as well as conference proceedings, across different disciplines. We will support this with two instruments. First, in close collaboration with editors and publishers of renowned and innovation-driven journals as well as conferences and their proceedings, we plan to create awareness of the approach in research communities through Author Guidelines as well as journal-specific examples of authors that have used the approach for their published papers. Such domain-specific examples arguably illustrate the benefits best.

Second, as a catalyzing instrument, we plan to develop a proposal for a German national project that aims to transfer the approach to numerous institutions that conduct research in diverse disciplines of the natural sciences, social sciences, and humanities. In close collaboration with selected research institutions, we envision collaborating with affiliated researchers by offering formal training and support to assure their research findings are produced and published also as machine-reusable expressions. Such a project could ensure that over the course of two to three years, hundreds of articles in dozens of journals produced by multiple publishers will have applied \emph{reborn} and would, thus, demonstrate its viability at scale in practice.

On a more technical level, we aim to address the current ORKG-specific syntax of the resulting supplementary data. The objective is to ensure high interoperability of the data produced and published in a distributed manner, and make the consumption of the data easier for aggregating systems other than ORKG. To achieve this, we are investigating the possibility to decouple data type specification from ORKG. More concretely, we envision aligning with the CNRI Data Type Registry (DTR, \href{https://typeregistry.org}{typeregistry.org}), specifically the instance operated by the ePIC Persistent Identifier Consortium. Data production would thus conform to data types specified by the DTR (\href{http://typeregistry.lab.pidconsortium.net/}{typeregistry.lab.pidconsortium.net}). To enable data collection, aggregating systems such as ORKG are then required to map the third-party data types to their own. In the ORKG, such a mapping can be supported with ORKG Templates. We argue that the decoupling of data type specification from any individual application or service will increase data interoperability and likely also adoption of the approach. As a further benefit, the ePIC DTR enforces a governance of specifications while, in its current implementation, anyone can edit ORKG Templates. A governance can ensure quality and stability, which is arguably essential for long-term viability of the system. 

Decoupling data type specification may also have positive repercussions on the development and maintenance of libraries for diverse (statistical) computing environments. Given the high dynamism of ORKG Templates, in their current implementation, the ORKG Python and R libraries dynamically generate a function-based native API on demand. This implementation has at least two downsides. First, the computing environment is dependent on internet connectivity at runtime. Second, the generation of the API at runtime is complicated. For the ePIC DTR with a governed and less dynamic type development, we no longer require a dynamic setup, but could instead generate the libraries in a static manner. As DTR types evolve, new versions of the libraries could be generated and deployed programmatically with little effort. Given the archiving of DTR types and libraries, the relationship between data produced with older libraries and the respective DTR type specification remains explicit.

To further improve the reusability of the machine-reusable expressions produced with \emph{reborn}, supporting terminological annotations of used vocabulary is an additional technical advancement. For instance, data input to and output from statistical data analysis are generally described with mostly ambiguous, hardly machine parsable parameter descriptions, e.g. ``MWD\_cor'', ``cc\_variant'', etc. A technique that enables researchers or (for the time being) a centralized data curation center to map such parameter descriptions to terms of community-agreed terminologies would further increase data reusability. This mapping could be supported directly in scripts, prior to passing data frames as values in templated supplementary data. With this, the published machine-reusable expressions of scientific knowledge would include term mappings, which can be exploited in data reuse to reduce ambiguity and improve automation, for instance in the conversion of quantity values with different units.

\begin{figure}[!ht]
\centering
\includegraphics[width=\linewidth]{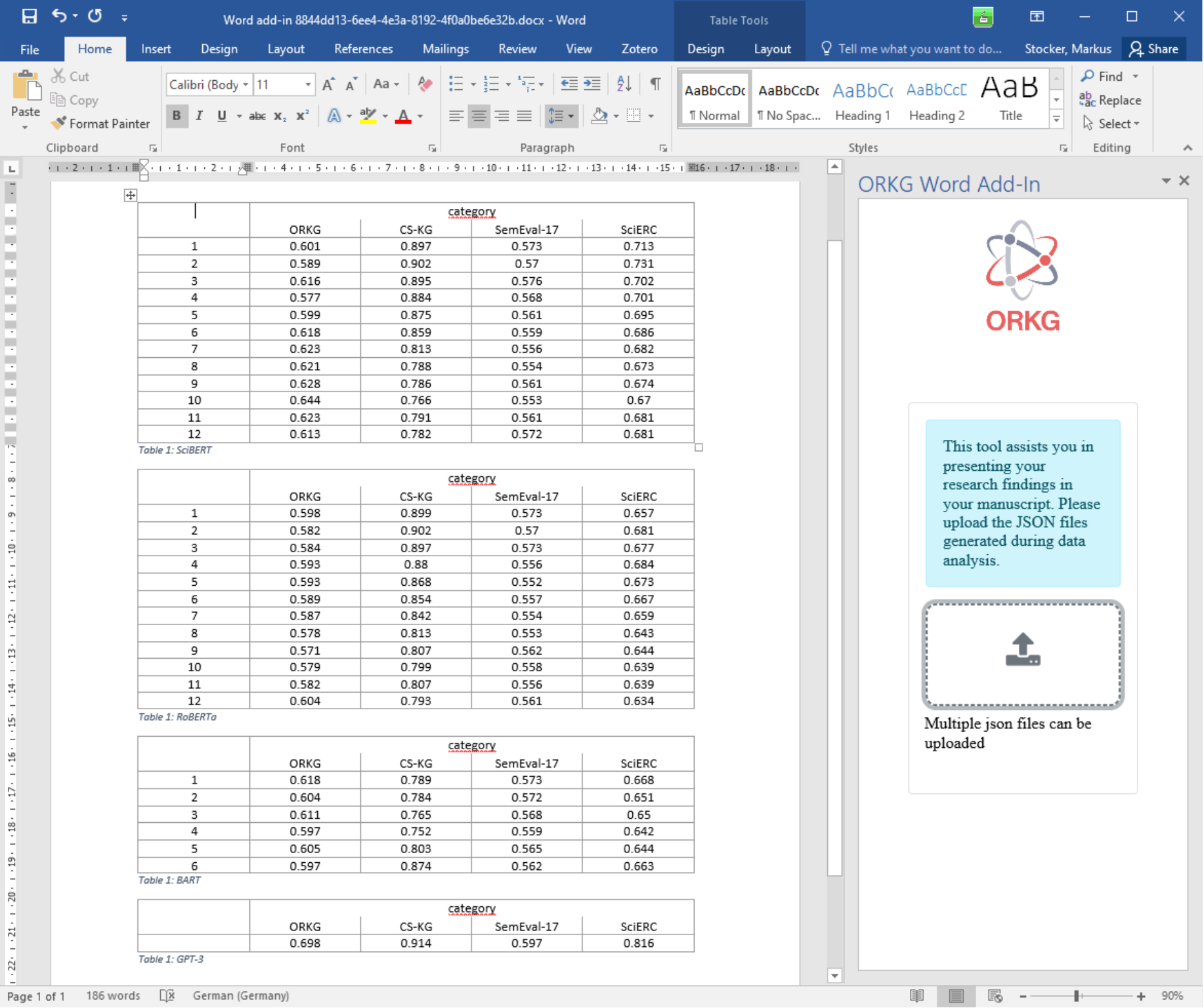}
\caption{ORKG Word Add-in display of the supplementary data in our use case in computer science for model performance for each dataset in the evaluation. Users provide the JSON-LD supplementary data produced in model performance evaluation and the Add-in automatically renders such TDMS-data as tables, one for each evaluated model.}
\label{fig:figure5}
\end{figure}

As a further avenue for future work, we are investigating the possibility of using the machine-reusable scientific knowledge produced in data analysis during the manuscript writing phase of the research lifecycle. Concretely, we are developing an MS Word Add-in that reads such data and assists manuscript writing by generating human-readable expressions for the encoded knowledge. As an example, if a researcher conducts a Student's t-test in Python and utilizes the proposed approach to produce a machine-reusable expression of the t-test, the resulting data are used by the Add-in to create a figure of the data input in the statistical test and a natural language sentence summarizing the test and resulting p-value. As another example, Figure \ref{fig:figure5} shows how the Add-in automatically displays the supplementary data of the use case in computer science presented earlier in a format suitable for a manuscript, namely as a set of tables summarizing model performance for each dataset in the evaluation. Naturally, these machine-generated elements are editable so that authors can manually refine them. The approach can also be implemented for manuscript editing software other than MS Word. The possibility of using machine-reusable scientific knowledge produced in data analysis during the writing of the related manuscript may be a strong incentive for researchers to adopt the proposed approach and benefit from producing the data directly in the lifecycle of their research.

\section*{Methods}

In this section, we present the proposed system in more detail. We begin with an architectural overview of the distributed system, its components, exchanged data, and interactions. Next, we present our reference implementation of the proposed system architecture.

\subsection*{Architectural overview}

\begin{figure}[!ht]
\centering
\includegraphics[width=\linewidth]{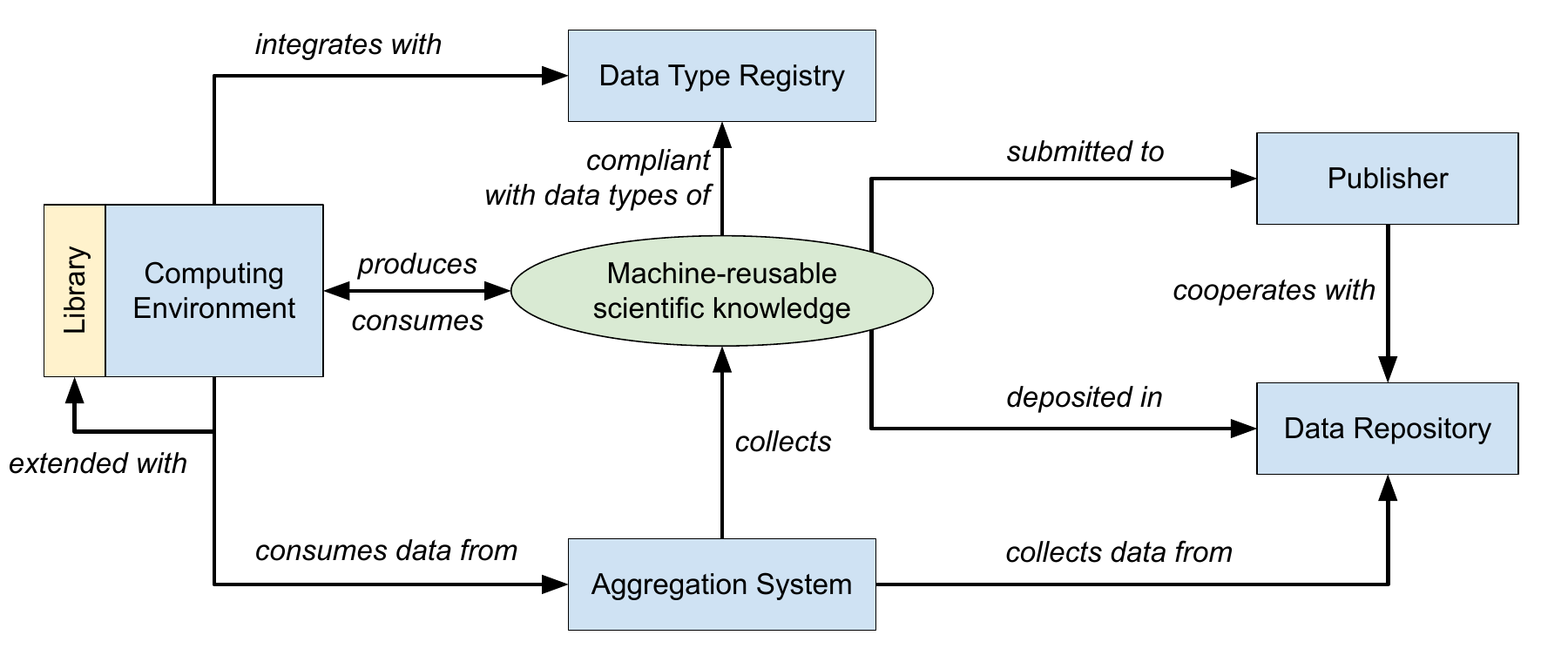}
\caption{Architectural overview of the proposed distributed system, its components and relations, including the (statistical) computing environment with specialized library, data type registry, publisher, data repository, and aggregation system; machine-reusable scientific knowledge as the primary exchanged data; and system-system as well as system-data relations.}
\label{fig:figure6}
\end{figure}

Figure \ref{fig:figure6} illustrates the architecture of the proposed distributed system. By supporting the production of scientific knowledge, the (statistical) Computing Environment is central to this system. Examples for such environments include the languages Python and R as well as the related systems Jupyter and R Studio. This environment is extended with a Library that provides users with functionality to produce and consume machine-reusable scientific knowledge. The computing environment may run on a personal computer or in the cloud. 

Researchers use (statistical) computing environments to implement the data analysis they require to produce the scientific knowledge expressed in articles (aka final data). It is within these environments that, supported by associated libraries, researchers ensure that scientific knowledge is also produced machine reusable. This production is guided by data types specified by a Data Type Registry. Ideally, the development of such data types is guided by existing conceptualizations, e.g. ontologies. For instance, the development of a knowledge type for Student's t-test is guided by a formalization of the concept in, e.g. the Ontology for Biomedical Investigations (OBI, \href{http://purl.obolibrary.org/obo/OBI\_0000739}{purl.obolibrary.org/obo/OBI\_00007399}).

By extending their scripts, authors can thus ensure that the (most salient) results are expressed in machine-reusable form. These data are submitted as supplementary data of the manuscript to the preferred journal or conference proceeding. Upon acceptance, the publisher ensures the publishing of the article and the deposition of the supplementary data in a data repository as well as article-data interlinking in DOI metadata. The supplementary data may also be deposited by a third-party organization that cooperates with publishers or by the authors using the preferred data repository (e.g. Dryad, Zenodo or similar). It is important for the data repository to support persistent URLs to individual files. A (persistent) identification only of the deposited collection as a whole is not sufficient. If deposited independently, authors are encouraged to ensure a high quality data publication that enables discovery via inverse interlinking, from supplementary data to published article, as well as reproducibility (if code snippets are interlinked). Scholarly infrastructures such as Crossref (\href{https://www.crossref.org/}{crossref.org}), DataCite (\href{https://datacite.org/}{datacite.org}), and OpenAIRE Graph (\href{https://graph.openaire.eu/}{graph.openaire.eu}) are meanwhile able to resolve direct and inverse links independently of which is created first.

Given an article DOI, Aggregation Systems collect deposited machine-reusable scientific knowledge and make these data accessible for efficient use. In addition to collecting data, aggregation systems provide value-added services, e.g. for scientific knowledge integration or visualization.

To support research, the machine-reusable scientific knowledge made available by aggregation systems can be consumed in (statistical) computing environments, thus closing the production-consumption loop. This interaction is particularly interesting for synthesis research, e.g. systematic reviews or meta-analyses.

\subsection*{Reference implementation}

Having presented the architecture of the proposed distributed system, we now turn to our reference implementation. In this implementation, the ORKG acts as the Aggregation System and also implements the Data Type Registry as a Template Registry. Furthermore, Jupyter and the Python language act as the (statistical) Computing Environment. Consequently, the ORKG Python library takes the role of the Library associated with the statistical computing environment. We describe the reference implementation by detailing the proposed distributed system along its main supported activities. Note that code listings shown here are subject to change as we further develop \emph{reborn}. Future changes will be reflected on our help center article available online at \href{https://orkg.org/help-center/article/47/reborn}{orkg.org/help-center/article/47/reborn}.

\paragraph{Creating templates.} The ORKG supports the creation of templates and maintains a template registry (\href{https://orkg.org/templates}{orkg.org/templates}). Users can thus search for existing templates and, assuming a suitable template is found, they can directly use templates to guide the production of machine-reusable scientific knowledge. Each template is identified by an ORKG Resource ID and is accessible via a unique URL. The main purpose of templates is to ensure that knowledge of the same type is represented in the same way. This supports knowledge integration. 

While any ORKG user can create templates, in general this is a task for experts, who, in the ORKG, organize themselves in the context of ORKG Observatories. Users who create templates should be experienced with developing schemata and know how to use ontologies to guide template creation. Ideally, the creation of templates is a collaborative effort and the resulting template specification reflects a community’s consensus. Only with such consensus can we ensure that templates are highly reusable in varying contexts.

Given that template creation is an activity driven by experts, most users should primarily use existing templates, and not be involved in their creation. We argue that the number of templates required to describe much of scientific knowledge is probably relatively small, especially if constrained to scientific knowledge that results in statistical data analysis. While there is a large number of statistical methods, it is possible that a relatively small number of methods is extensively used in research. Student's t-test, ANOVA, and linear regressions are good examples. Hence, we expect to see the need for template creation diminish as users will increasingly find the required templates in the registry.

\paragraph{Integration of templates in data analysis.} In the ORKG, templates were originally designed and developed to support the manual description of research contributions using the ORKG frontend at \href{https://orkg.org}{orkg.org}. In the context of the present work, we have extended the application of templates in data analysis implemented in computing environments. This is supported by corresponding libraries: for a Python-based statistical computing environment, it is the ORKG Python library (\href{https://orkg.readthedocs.io}{orkg.readthedocs.io}). Among other features, this library supports the integration of ORKG Templates in Python scripts that implement statistical data analysis activities that underlie the results published in articles. A corresponding library is also available for environments employing the R language.

Listing \ref{lst:listing1} illustrates how ORKG Templates are integrated into Python scripts, for simplicity exemplified on a Student's t-test (e.g. petal length for two Iris species) using the well-known Iris dataset. Having imported libraries, including the ORKG library, at line 6, we first load the Iris dataset, which is then also processed (not shown). At line 8, we perform the t-test and obtain the computed p-value on the following line. The integration of ORKG Templates into scripts is shown on the following lines. At line 11, we connect to the ORKG production system at orkg.org. At line 12, we materialize a template by its ORKG Resource ID, here R12002, which is the ORKG Template for Student’s t-tests. Finally, at line 13, we create an object that provides access to the materialized template.

The materialization step is interesting because it dynamically creates a function-based native API based on the template specification. As ORKG Templates can be nested, the materialization step recursively materializes all templates that are nested in the template addressed by ID.

Lines 15-21 illustrate the use of the dynamically created API to instantiate the template with data; here, to describe the conducted Student's t-test. Following the template, a Student's t-test has a label, a dependent variable, an input dataset, and an output p-value. Notably, the input dataset is a Python data frame, and we can simply pass the variable already created in reading/processing the input data, e.g. from a local CSV file. Moreover, the p-value results from the statistical hypothesis test and is also passed as a variable. Finally, at line 22, we instruct the library to write the data expressing the conducted t-test in machine-reusable form to a file in JSON-LD format.

\paragraph{Depositing supplementary data.} The JSON-LD data files produced in data analysis are first stored in the user's computing environment (e.g. on their personal computer). These data files and related data files, including script snippets or figures, are later submitted as supplementary data together with the manuscript to the co-authors' preferred journal for manuscript review. At this point, the supplementary data enters the review and publication process, and journals or conference chairs may consider using the supplementary data to support the review process. While authors and publishers may publish the supplementary data themselves, in our reference implementation we use the TIB Leibniz Data Manager \cite{beer2022ldm} as a default data repository. Centralizing the publication of the supplementary data may be (at least for the time being) the most efficient approach, as TIB has the expertise to ensure high quality publishing in terms of identification, persistence, interlinking, and reproducibility.

\paragraph{Interlinking supplementary data in metadata.} To enable DOI-based discovery of harvestable supplementary data by machines, we support the interlinking of assets in DOI metadata. We follow two approaches: data-to-article and (optionally) article-to-data interlinking. Listing \ref{lst:listing2} exemplifies data-to-article interlinking, which is the primary approach. We interlink the article DOI in dataset DOI metadata using the \texttt{IsSupplementTo} relation. Furthermore, we use the \texttt{HasPart} relation to link explicitly to JSON-LD supplementary data. In addition, publishers may (optionally) implement article-to-data interlinking with the \texttt{is-supplemented-by} (or similar) relation in Crossref DOI metadata according to the pattern illustrated in Listing \ref{lst:listing3}. Listing \ref{lst:listing4} exemplifies the discovery of supplementary JSON-LD data given the article DOI using the DataCite REST API.

\paragraph{Collecting supplementary data in aggregation systems.} Given supplementary data interlinked in DOI metadata about the article, aggregation systems only need the article DOI to discover harvestable data. For the ORKG, its Python library also supports DOI- and directory-based harvesting of data into the ORKG. Listing \ref{lst:listing5} illustrates this feature with an example. Python-based harvesting, e.g. on the command line, is particularly useful for work published in articles not identified by a DOI or to ingest into ORKG data that are supplementary to manuscripts in review. Authors or publishers can perform such harvesting, and, in the ORKG, harvesting may be into sandbox or production systems. Note that supplementary data produced with templates on ORKG sandbox (\href{https://sandbox.orkg.org}{sandbox.orkg.org}) are in general not compatible with ORKG production (\href{https://orkg.org}{orkg.org}). This is because templates on sandbox and on production may have the same structure, but the properties and classes used in the respective templates have different IDs. This issue can be addressed by decoupling the templating from the ORKG, as proposed using, e.g. the ePIC Data Type Registry (see Future Work).

\paragraph{Integrating aggregation systems in data analysis.} Closing the loop for the production and consumption of machine-reusable scientific knowledge in statistical computing environments, the ORKG Python and R libraries support the retrieval of ORKG tabular data into data frames native to the respective languages. This enables the frictionless machine-based reuse of scientific knowledge, e.g. in synthesis research. Listing \ref{lst:listing6} illustrates this feature with an example. As this work focuses on the production and publication of machine-reusable scientific knowledge, we do not further explore the possibilities of reuse here and defer this aspect to future articles.

\section*{Data and code availability}


We applied \emph{reborn} to the present article and expressed the most salient aspects of the three use cases in structured form. Readers may access the ORKG Paper for the present work at \href{https://orkg.org/paper/R691379}{orkg.org/paper/R691379}. Using the ORKG Comparison feature, it is possible to juxtapose the three use cases in a tabular form (\href{https://orkg.org/comparison?contributions=R691381,R691401,R691391}{orkg.org/comparison?contributions=R691381,R691401,R691391}). The ORKG and its components are released Open Source under the MIT license. The source code for the ORKG and its components, including the Python and R libraries, are available at \href{https://gitlab.com/TIBHannover/orkg/}{gitlab.com/TIBHannover/orkg}. The data constituting the use cases presented in this article are Open Data published by the ORKG under the CC0 1.0 Universal license. In addition to the ORKG Frontend, the data can be accessed via numerous APIs, including REST and SPARQL. Information about ORKG data access is available at \href{https://orkg.org/data}{orkg.org/data}. Supplementary data depositions in the TIB Leibniz Data Manager are published under CC0 1.0 Universal license. This includes code snippets used to compute the research findings and resulting figures that are interlinked in the respective ORKG research contributions.

\bibliography{sample}


\section*{Acknowledgements}


The authors greatly appreciate the support provided by Norman Gentsch and Ricardo Perez-Alvarez, first authors of the original works constituting the respective use cases underlying this article. We thank our colleagues Olga Lezhnina for supporting the ORKG R package, Lars Vogt for supporting the conceptualization of templates, Qurat-ul Ain Aftab for maintaining the ORKG Word Add-in, and Manuel Prinz for related ORKG backend developments. Finally, we thank Chris Mavergames and Wiley as well as Johannes Wagner and Copernicus Publications for their support of this work. Parts of the work described in this article have been co-funded by the European Research Council (ERC) project ScienceGRAPH (GA: 819536), the German Research Foundation (DFG) project NFDI4DS (PN: 460234259), and the Leibniz University Hannover (LUH) Flexible Funds (PN: FlexFunds-2022-05). This research has also been supported by the Leibniz Research Network ``Earth \& Societies'' and the Leibniz-Lab ``Systemic Sustainability'' funded by the Leibniz Association.

\section*{Author contributions statement}


Author contributions according to the CRediT taxonomy: Conceptualization (MS); Data curation (MS, LS, MA, FT); Formal Analysis (N/A); Funding acquisition (MS, LS); Investigation (MS, LS, MA, FT); Methodology (MS, LS, OL, KEF, MH, AO, MYJ); Project administration (MS, LS); Resources (MS, LS, MA, FT); Software (MS, MA, OL, FT, KEF, MH, AO, MYJ); Supervision (MS, LS); Validation (MS, LS, MA, FT); Visualization (MS); Writing – original draft (MS, LS); Writing – review \& editing (MS, LS, MA, OL, FT, KEF, MH, AO, MYJ). The authors did not use generative AI technology such as AI chatbots to assist activities related to this research, including the writing of this article.

\section*{Competing interests}


The authors declare no competing interests.

\newpage

\nolinenumbers

\section*{Figures \& Tables}





\newpage

\newpage

\newpage

\newpage

\newpage

\newpage

\begin{table}[!ht]
\centering
\begin{tabular}{|l|c|c|c|c|c|c|}
\hline
Approach & Accuracy & Richness & Simplicity & Scalability & Coverage & Legacy \\
\hline
Post-publication extraction, manually & \cmark & \xmark & \cmark & \xmark & \cmark & \cmark \\
\hline
Post-publication extraction, automatically & \xmark & \xmark\xmark & \xmark\xmark & \cmark\cmark & \cmark\cmark & \cmark\cmark \\
\hline
Pre-publication production & \cmark\cmark & \cmark\cmark & \cmark\cmark & \xmark & \cmark & \xmark \\
\hline
\end{tabular}
\caption{Overview of the pros and cons of classical post-publication extraction in contrast to the proposed pre-publication production, compared along important dimensions.}
\label{tbl:table1}
\end{table}

\newpage

\begin{lstlisting}[caption={Integration of templates in a Python script, here shown for the Student's t-test template with resource ID R12002 exemplified with the Iris dataset. Note that for the sake of simplicity we illustrate this step with a popular dataset and statistical method, rather than with our running example in soil science. Furthermore, we omit some details on Iris dataset processing.}, frame=tb, captionpos=t, basicstyle=\ttfamily\small, numbers=left, label=lst:listing1]
from orkg import ORKG
from sklearn import datasets
from scipy.stats import ttest_ind

# Load Iris dataset and select x,y vectors (versicolor and virginica petal length)
iris = datasets.load_iris()
 
tt = ttest_ind(x, y, equal_var=False)
pvalue = tt.pvalue

orkg = ORKG(host="https://orkg.org")
orkg.templates.materialize_template("R12002") 
tp = orkg.templates

tp.students_ttest(
	label="Statistically significant hypothesis test for petal length of iris species",
	has_dependent_variable="http://purl.obolibrary.org/obo/TO_0002605",
	has_specified_input=(iris, "the Iris dataset"),
	has_specified_output=tp.pvalue("the p-value",
		tp.scalar_value_specification(pvalue)
	)
).serialize_to_file("article.contribution.1.json", format="json-ld")
\end{lstlisting}

\newpage

\begin{lstlisting}[caption={Data-to-article interlinking in DataCite DOI metadata using the \texttt{IsSupplementTo} and \texttt{HasPart} relations (see also: \protect\href{https://api.datacite.org/dois/10.57702/yztrbsd4}{api.datacite.org/dois/10.57702/yztrbsd4}).}, frame=tb, captionpos=t, basicstyle=\ttfamily\small, numbers=left, label=lst:listing2]
"relatedIdentifiers": [
  {
    "relationType": "IsSupplementTo",
    "relatedIdentifier": "10.5194/soil-10-139-2024",
    "resourceTypeGeneral":"JournalArticle",
    "relatedIdentifierType":"DOI",
  },
  {
    "relationType":"HasPart",
    "relatedIdentifier":"https://service.tib.eu/.../contribution-1.json",
    "resourceTypeGeneral":"Dataset",
    "relatedIdentifierType":"URL"
  },
  ...
]
\end{lstlisting}

\newpage

\begin{lstlisting}[caption={Optional article-to-data interlinking in Crossref DOI metadata using the \texttt{is-supplemented-by} relation.}, frame=tb, captionpos=t, basicstyle=\ttfamily\small, numbers=left, label=lst:listing3]
"relation": {
  "is-supplemented-by": [
    {  
      "id-type": "doi",
      "id": "10.57702/yztrbsd4"
    }
  ]
}
\end{lstlisting}

\newpage

\begin{lstlisting}[caption={Discovery of supplementary data by article DOI using the DataCite REST API.}, frame=tb, captionpos=t, basicstyle=\ttfamily\small, numbers=left, label=lst:listing4]
https://api.datacite.org/dois?query=
  relatedIdentifiers.relatedIdentifier:10.5194/soil-10-139-2024 
  AND relatedIdentifiers.relationType:IsSupplementTo
\end{lstlisting}

\newpage

\begin{lstlisting}[caption={Harvesting supplementary data into the ORKG with Python using its DOI and directory-based harvesting approaches.}, frame=tb, captionpos=t, basicstyle=\ttfamily\small, numbers=left, label=lst:listing5]
from orkg import ORKG

orkg.harvesters.doi_harvest(
  doi="https://doi.org/10.5194/soil-10-139-2024",
  orkg_rf="Soil Science"
)

orkg.harvesters.directory_harvest(
  directory="data",
  research_field="Soil Science",
  title="Cover crops improve soil structure and change organic carbon distribution 
         in macroaggregate fractions",
  authors=["Norman Gentsch", "Florin Laura Riechers", "Jens Boy", 
           "Dörte Schwenecker", "Ulf Feuerstein", "Diana Heuermann", 
           "Georg Guggenberger"],
  publication_year=2024,
  publication_month=9,
  published_in="SOIL"
)
\end{lstlisting}

\newpage

\begin{lstlisting}[caption={Retrieving ORKG tabular data as a data frame in Python.}, frame=tb, captionpos=t, basicstyle=\ttfamily\small, numbers=left, label=lst:listing6]
from orkg import ORKG
 
orkg = ORKG(host="https://orkg.org")
df = orkg.resources.by_id(id="R662664").as_dataframe()
df.head()
\end{lstlisting}

\end{document}